# Denoising diffusion-based synthetic generation of three-dimensional (3D) anisotropic microstructures from two-dimensional (2D) micrographs

Kang-Hyun Lee[1], and Gun Jin Yun[1,2,*]

[1]*Department of Aerospace Engineering, Seoul National University, Gwanak-gu Gwanak-ro 1 Seoul 08826, South Korea*
[2]*Institute of Advanced Aerospace Technology, Seoul National University, Gwanak-gu Gwanak-ro 1, Seoul 08826, South Korea*

**Abstract**

Integrated computational materials engineering (ICME) has significantly enhanced the systemic analysis of the relationship between microstructure and material properties, paving the way for the development of high-performance materials. However, analyzing microstructure-sensitive material behavior remains challenging due to the scarcity of three-dimensional (3D) microstructure datasets. Moreover, this challenge is amplified if the microstructure is anisotropic, as this results in anisotropic material properties as well. In this paper, we present a framework for reconstruction of anisotropic microstructures solely based on two-dimensional (2D) micrographs using conditional diffusion-based generative models (DGMs). The proposed framework involves spatial connection of multiple 2D conditional DGMs, each trained to generate 2D microstructure samples for three different orthogonal planes. The connected multiple reverse diffusion processes then enable effective modeling of a Markov chain for transforming noise into a 3D microstructure sample. Furthermore, a modified harmonized sampling is employed to enhance the sample quality while preserving the spatial connection between the slices of anisotropic microstructure samples in 3D space. To validate the proposed framework, the 2D-to-3D reconstructed anisotropic microstructure samples are evaluated in terms of both the spatial correlation function and the physical material behavior. The results demonstrate that the framework is capable of reproducing not only the statistical distribution of material phases but also the material properties in 3D space. This highlights the potential application of the proposed 2D-to-3D reconstruction framework in establishing microstructure-property linkages, which could aid high-throughput material design for future studies

**Keywords:** Denoising diffusion; Three-dimensional (3D) microstructure reconstruction; Generative models; Multiphase/anisotropic microstructures;

## 1. Introduction

In recent decades, establishing the process-microstructure-property linkage has become crucial for material design, especially with the aid of computational micromechanics [1-3] and the integrated computational materials engineering (ICME) framework [4-6]. For instance, multiscale computational analysis that employ the concept of the representative volume element (RVE) have been widely utilized to analyze the representative material response under specific boundary conditions [2, 7]. In particular, computational homogenization methods accompanying with finite element analysis (FEA) based on the asymptotic homogenization theory [8, 9], have been used to analyze the macroscopic properties of representative microstructures (i.e., RVEs) for various types of microstructural materials including particulate or fibrous composites [10-14], multi-phase polycrystalline metals [15, 16], metal matrix composites [17, 18], and lattice materials [19-21]. In general, these methods assume that the behavior of a heterogeneous material can be described by an RVE that is periodic throughout the material of interest. If an RVE is properly modeled for the subsequent computational homogenization analysis with periodic boundary conditions (PBC), and the macro-structure is sufficiently large, accurate solutions for the homogenized material properties (i.e., effective material properties) can be obtained. The homogenized properties of heterogenous materials can also be acquired based on the fast Fourier transform (FFT) [22, 23], which avoids the time-consuming FEA for computing the material response under macroscopic loading.

---
* Corresponding author, Professor, Department of Mechanical & Aerospace Engineering, Seoul National University, Gwanak-ro 1 Gwanak-gu, Seoul, 08826, South Korea, Tel)+82-2-880-8302, Email) gunjin.yun@snu.ac.kr

Meanwhile, the recently proposed deep learning (DL) models [24-28] that link microstructure to material properties are gaining significant attention, due to their remarkably lower computational cost compared to conventional computational homogenization methods. For instance, Rao and Liu developed a three-dimensional convolutional neural network (3D-CNN) for homogenization of heterogeneous materials with random spherical inclusions [24]. Their results showed that after training the 3D-CNN using the training data pairs (i.e., microstructure RVEs and anisotropic material properties), the model could accurately estimate the anisotropic elastic material properties with a maximum prediction error of up to 0.5%. Eidel [27] proposed universal 3D-CNNs to predict the macroscale elastic stiffness based on given RVEs of random heterogeneous materials, with the intent of replacing FEA-based computational homogenization. In addition, the term 'universal' in their study refers to the capability of their 3D-CNNs, which can handle various types of microstructures, arbitrary phase fractions, and wide ranges of elastic moduli of the constituent phases. Cheng and Wagner [28] proposed a fully convolutional network-based framework called RVE-net, designed to derive material response from given microstructural geometry while offering the capability for automatic defect characterization. Their framework also employs the residual of the governing equations and boundary conditions in the form of a loss function for incorporating specific physics into neural network models. In summary, many studies have sought to establish a connection between microstructures and material properties. Given sufficient information about microstructural geometries, the analysis of material response and behavior under specific boundary conditions can be conducted using the introduced methods.

On the other hand, acquiring microstructure data, such as two-dimensional (2D) micrographs from optical microscopy and scanning electron microscopy (SEM) [29], requires time and expense for experimentation, and this becomes even more challenging in the case of 3D microstructure data. Since material behavior (e.g., elastic and plastic behavior) of heterogenous material is highly sensitive to the 3D microstructural geometry [30-32], it is imperative to analyze the geometrical features in 3D space to comprehend the material response under specific boundary conditions. In this regard, micro-computed tomography (micro-CT) has been utilized to capture the 3D details of material microstructures [33-40]. Micro-CT enables the visualization of a material's internal structure in a non-destructive manner, which also facilitates the modeling of RVEs for computational analysis of material behavior [34, 35, 37-39]. However, micro-CT scans necessitate careful processing of the raw data, and the selected filter parameters and threshold values have a profound impact on the resulting microstructural geometry, which in turn affects the predicted material response [41-43]. Furthermore, performing micro-CT scans for specific materials and length scales can be either cost-prohibitive or technically impractical due to resolution limits and material sensitivity [33]. To resolve this problem, various microstructure characterization and reconstruction (MCR) algorithms have been developed to characterize the 2D microstructures with certain microstructural descriptors (e.g., volume fraction, n-point correlation functions, surface area, shape index, equivalent particle radius, etc.) and to reconstruct equivalent 3D microstructure samples [40, 44-47]. After characterizing a representative 2D microstructure sample through specific descriptors (i.e., target descriptors), optimization methods (e.g., stochastic algorithms and gradient-based optimization) can be employed to create new equivalent 3D samples by minimizing the difference between the descriptors of the current sample and the target descriptors [40, 46, 48, 49].

However, the aforementioned MCR methods require careful selection of descriptors to ensure the quality of the generated samples. Since the spatial distribution and morphological characteristics of microstructures vary depending on the material of interest, proper characterization of microstructures is crucial for generating equivalent microstructure samples [45, 46]. Meanwhile, a considerable amount of research has focused on using deep generative models to reconstruct microstructures without relying on microstructural descriptors [45, 50-60]. Several studies have demonstrated that microstructure reconstruction using generative adversarial networks (GANs) [50-52] or variational autoencoders (VAEs) [53, 55, 56] can reconstruct microstructure samples comparable to the original training micrographs. In particular, Seibert and Cooper [52] also introduced a novel GAN architecture called 'SliceGAN' to reconstruct 3D microstructures from 2D micrographs for a variety of isotropic and anisotropic microstructures. Additionally, the DL models can be considered as the potential methods for developing an universal microstructure reconstruction framework, as they characterize microstructure geometry in latent space representations instead of using specific descriptors tailored for distinct microstructures [61, 62]. Nonetheless, there have been reports of significant issues related to generative models that obstruct their extensive use for sampling novel data, such as the generation of blurred samples by VAE-based models [63, 64] and the occurrence of mode collapse resulting from the adversarial loss function in GANs [65-67]. These problems are significant because, as previously mentioned, the behavior of heterogeneous materials is highly sensitive to microstructural geometry. Moreover, 2D-to-3D microstructure reconstruction using deep generative models has not been studied extensively due to their 'black box' nature and the limited interpretability of the latent space. The challenge of reconstructing anisotropic microstructures, rather than isotropic microstructures, also needs to be addressed further.

Recently, diffusion-based generative models (DGMs) have attracted considerable attention for being a promising generative model of the next generation, which is less susceptible to mode collapse and unstable training process [68-71]. In particular, denoising diffusion probabilistic models (DDPMs), which are a parameterized version of DGMs consisting of a reverse Markovian chains for denoising a data structure from Gaussian noise, have surpassed GANs in the context of image generation tasks [72]. In addition, DGMs have shown outstanding generative performance across various fields such as natural language processing [73, 74], computer vision [75-77], multimodal modeling [78-80] and medical image processing [81, 82]. Several recent studies have also demonstrated that 2D microstructure samples can be generated using DGMs trained with 2D micrograph datasets, whether synthesized or experimentally obtained [57, 83]. For instance, Lee and Yun [57] demonstrated that DGMs can generate various types of equivalent 2D microstructure samples that have spatial distributions of material phases similar to those in the 2D training micrographs. Furthermore, Düreth et al [83] showed that conditional DGMs can be used to reconstruct various types of real-world microstructure data, validating their results using descriptor-based error metrics and the Fréchet Inception Distance (FID). Nonetheless, these methods focus on generating 2D samples with models trained with 2D images. Indeed, it was not until Lee and Yun [84] introduced a novel framework known as 'Micro3Diff' that the reconstruction of 3D microstructures from 2D micrographs with DGMs was presented. They demonstrated that their framework is capable of generating 3D microstructure samples using 2D-DGMs (i.e., DGMs trained only with 2D images) by leveraging the latent variables of DGMs using a multi-plane denoising diffusion and harmonized sampling approach. Nonetheless, the scope of their framework was restricted to generating binary microstructures, which are isotropic within a 3D space. To fully exploit DGMs for the reconstruction of microstructures, a methodology must be developed to reconstruct various types of microstructures, including anisotropic and multiphase microstructures.

In this study, a conditional DGM-based microstructure reconstruction framework is proposed for generating 3D anisotropic microstructures solely based on 2D micrographs. The 2D-to-3D reconstruction of multiphase microstructure is also addressed. First, the proposed framework involves the training of conditional 2D-DGMs to generate 2D microstructure samples at three orthogonal planes, which is equivalent to sampling images from the underlying conditional distributions of the microstructure data. Then, the trained 2D-DGMs are used for generating 3D anisotropic microstructures with simultaneous denoising process at three orthogonal planes according to the embedded conditions for guidance. In other words, several spatially connected denoising processes occur within a single sampling process, effectively modeling a Markov chain to generate 3D anisotropic structures. In the following section, the detailed methodology for building the overall framework is introduced as well as the background knowledge regarding DGMs and 2D-to-3D reconstruction of data. Then, we assess the performance of the proposed framework with different types of microstructures to validate the proposed methodology both qualitatively and quantitatively.

**2. Methodology**

2.1 Denoising diffusion probabilistic models (DDPMs)

Among the different formulations of DGMs including the score-based generative models (SGMs) and stochastic differential equations (SDEs), DDPMs are widely used for implementing DGMs [68, 72, 73, 78-80, 85]. To build DGMs for generating microstructure samples, the formulation of DDPMs was employed in this study. In addition, the formulations of DGMs can be generalized into SDEs with a continuous time variable, as elaborated in detail in our previous work [84] and other references [70, 71].

In general, DDPMs involve two types of processes: (1) the forward diffusion process, which gradually adds noise to the samples, and (2) the reverse diffusion process, which denoises the noisy samples to restore the original structures (Figure 1). With a pre-defined number of time steps $T$ (which usually numbers in thousands [72]), an original sample $\mathbf{x}_0$ can be noised following the discrete time sequence $t = [1, 2, 3, ..., T]$ which is known as the forward process. The noised sample at each time step $t$ is denoted as $\mathbf{x}_t$, which serves as the latent variable of DDPMs. According to Ho et al. [68], the forward process $p(\mathbf{x}_t|\mathbf{x}_{t-1})$ of DPDMs can be defined as

$$p(\mathbf{x}_t \mid \mathbf{x}_{t-1}) = N(\mathbf{x}_t; \sqrt{1 - \beta_t}\mathbf{x}_{t-1}, \beta_t \mathbf{I}) \qquad \text{Eq. 1}$$

where $\beta_t \in (0,1)$ is the noise schedule parameter at time $t$. It is also worth noting that the iterative noising process for a given time sequence can be simply replaced by the following equation:

$$p(\mathbf{x}_t \mid \mathbf{x}_0) = N(\mathbf{x}_t; \sqrt{\bar{\alpha}}\mathbf{x}_0, (1-\bar{\alpha}_t)\mathbf{I}) \qquad \text{Eq. 2}$$

where

$$\alpha_t = 1 - \beta_t \qquad \text{Eq. 3}$$

$$\bar{\alpha}_t = \prod_{i=0}^{t} \alpha_i \qquad \text{Eq. 4}$$

In particular, the sequence of noising process $p(\mathbf{x}_t|\mathbf{x}_{t-1})$ through the predefined time sequence formulates a Markovian chain, gradually transforming a sample to Gaussian noise. Since the forward process (Eq. 1) has the form of a conditional Gaussian distribution, a neural network model $p_\theta(\mathbf{x}_{t-1} \mid \mathbf{x}_t)$ for denoising process (i.e., reverse diffusion process) can be built with the following definition:

$$p_\theta(\mathbf{x}_{t-1}|\mathbf{x}_t) = N(\mathbf{x}_{t-1}; \mathbf{m}_\theta(\mathbf{x}_t, t), \mathbf{C}_\theta(\mathbf{x}_t, t)) \qquad \text{Eq. 5}$$

where $\mathbf{m}_\theta(\mathbf{x}_t, t)$ is the predicted mean function and $\mathbf{C}_\theta(\mathbf{x}_t, t)$ is the predicted covariance of the denoised sample $\mathbf{x}_{t-1}$. However, Ho et al. [68] demonstrated that $\mathbf{C}_\theta$ can be considered to a constant $\mathbf{C}_\theta(\mathbf{x}_t, t) = \beta_t \mathbf{I}$, and showed that focusing solely on learning the mean function results in an enhancement of sample quality. Thus, the loss function $L$ for training the model $p_\theta(\mathbf{x}_{t-1}|\mathbf{x}_t)$ can be written as

$$L = \mathbb{E}_p[\lambda(t)\|\mathbf{m}_\theta(\mathbf{x}_t, t) - \mathbf{m}_t(\mathbf{x}_t, \mathbf{x}_0)\|^2] \qquad \text{Eq. 6}$$

where $\mathbf{m}_t(\mathbf{x}_t, \mathbf{x}_0)$ is the mean function of noised sample $\mathbf{x}_t$ and $\lambda(t)$ denotes the positive weighting function. Moreover, rather than training a model to predict the mean of the denoised sample as Eq. 7, we can train a model $\boldsymbol{\epsilon}_\theta(\mathbf{x}_t, t)$ to estimate the noise component at $t$ by reparameterization. Then, we can denoise a sample $\mathbf{x}_t$ using the model $\boldsymbol{\epsilon}_\theta(\mathbf{x}_t, t)$ based on the definition of the forward process (Eq. 2) as

$$\mathbf{m}_\theta(\mathbf{x}_t, \mathbf{y}, t) = \frac{1}{\sqrt{\alpha_t}}\left(\mathbf{x}_t - \frac{1-\alpha_t}{\sqrt{1-\bar{\alpha}_t}}\boldsymbol{\epsilon}_\theta(\mathbf{x}_t, t)\right) \qquad \text{Eq. 8}$$

In addition, optimizing the loss function (Eq. 6) is equivalent to variational inference with a variational lower bound for learning the actual distribution of $p(\mathbf{x}_{t-1}|\mathbf{x}_t)$, as demonstrated in [68].

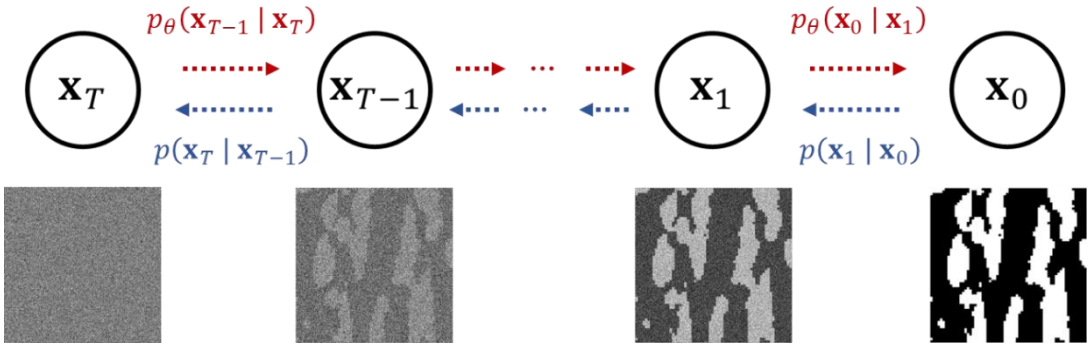

**Figure 1.** Schematic of forward and reverse diffusion processes (i.e., noising and denoising processes) in DDPMs within the discrete time sequence

## 2.2 Classifier-free guidance for conditional sampling

It is worth noting that the aforementioned forward and reverse processes of DDPMs (Eq. 1 and Eq. 5) are conditional only on the time step and the previous samples following the forward or reverse sequence. Since the data itself does not contain any conditions, the overall generation process itself is unconditional. To sample data from conditional distributions using DDPMs, it is required to add a conditional signal for guiding the reverse diffusion process at each time step $t$. In this regard, the classifier-free guidance [86] was employed for conditional sampling, which is necessary for generating anisotropic 3D microstructures (section 2.3) by simultaneously denoising the 2D samples at different planes (i.e., xy, xz, and yz planes). In other words, we first need conditional 2D-DGMs that can generate 2D microstructure samples at three orthogonal planes. According to Ho and Salimans [86], a single DGM can be trained for conditional sampling by randomly dropping the conditional signal with the following parameterized equation:

$$\hat{\boldsymbol{\epsilon}}(\mathbf{x}_t, \mathbf{y}, t) = \omega \boldsymbol{\epsilon}_\theta(\mathbf{x}_t, \mathbf{y}, t) + (1 - \omega)\boldsymbol{\epsilon}_\theta(\mathbf{x}_t, t) \qquad \text{Eq. 9}$$

where $\hat{\boldsymbol{\epsilon}}(\mathbf{x}_t, \mathbf{y}, t)$ is the resultant estimation of the noise component, $\boldsymbol{\epsilon}_\theta(\mathbf{x}_t, \mathbf{y}, t)$ and $\boldsymbol{\epsilon}_\theta(\mathbf{x}_t, t)$ are the conditional and unconditional estimations of the noise component at time step $t$, respectively. In addition, $\mathbf{y}$ denotes the input condition which represents a particular plane (xy or xz or yz-plane) for obtaining anisotropic microstructure samples in this study. Eq. 9 shows that the estimation becomes totally unconditional when $\omega = 0$ and totally conditional when $\omega = 1$. By randomly omitting the conditional estimation term during the training of DDPMs, the models can incorporate $\mathbf{y}$ during the reverse diffusion process. Based on the previous study [86], the probability of omitting the conditional signal for guidance was set to be 10%, optimizing both the unconditional and conditional objectives for DDPMs.

## 2.3 2D-to-3D reconstruction of anisotropic structures with DGMs

### 2.3.1 Problem formulation

If a 3D microstructure is decomposed into three sets of 2D micrographs corresponding to three orthogonal planes (Figure 2), the 2D data from the different planes ($\mathbf{x}_{xy}$, $\mathbf{x}_{xz}$ and $\mathbf{x}_{yz}$) can initially be regarded as samples from an unconditional data distribution (i.e., $p(\mathbf{x}_{xy}), p(\mathbf{x}_{xz})$ and $p(\mathbf{x}_{yz})$). In addition, this scenario is akin to real-world situations where 2D micrographs are typically collected experimentally without regard to the internal 3D microstructure. However, since the 2D slices are spatially connected, forming a 3D microstructural geometry, they are related and conditional on the data from different planes. For instance, the distribution of 2D micrographs at three different planes can be written as follows, considering an anisotropic 3D microstructure:

$$\mathbf{x}_{xy} \sim p(\mathbf{x}_{xy}|\mathbf{x}_{xz}, \mathbf{x}_{yz}) \qquad \text{Eq. 10}$$

$$\mathbf{x}_{yz} \sim p(\mathbf{x}_{yz}|\mathbf{x}_{xy}, \mathbf{x}_{xz}) \qquad \text{Eq. 11}$$

$$\mathbf{x}_{xz} \sim p(\mathbf{x}_{xz}|\mathbf{x}_{xy}, \mathbf{x}_{yz}) \qquad \text{Eq. 12}$$

This implies that we need a labeled dataset of 2D micrographs beforehand to build generative models than can sample data from the estimated distributions for Eq. 10 - Eq. 12. In other words, we need 2D samples that are spatially connected and assembled into data pairs for training generative models, indicating that a 3D dataset is necessary to generate a 3D structure. Since this approach does not qualify as 2D-to-3D reconstruction, an alternative method is required for employing generative models to create 3D samples from 2D samples

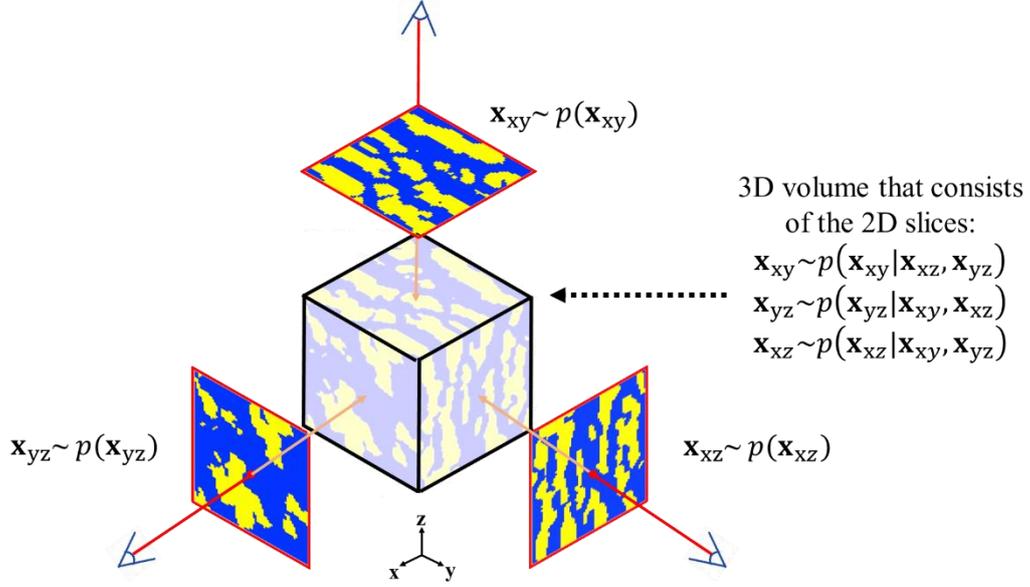

**Figure 2.** Unconditional/conditional distributions of 2D slices in a 3D volume in terms of three different viewpoints (xy/xz/yz-planes)

2.3.2 Multi-plane denoising diffusion for anisotropic microstructures

To enable 2D-to-3D reconstruction of anisotropic microstructures, this study proposes a conditional DGM-based framework relying solely on 2D micrographs (Figure 3). Compared to the generative models such as VAEs [53, 55, 56] and GANs [50-52] that use a single forward pass to generate a novel sample, DGMs use multiple reverse diffusion steps to transform noise into the original data structure. By leveraging the multiple diffusion steps of DGMs and the latent variables (i.e., $\mathbf{x}_t$), Lee and Yun [84] demonstrated a framework called 'Micro3Diff' for 2D-to-3D reconstruction of isotropic microstructures based on the concept of multi-plane denoising diffusion. In this study, the capability of DGMs is extended to address 2D-to-3D reconstruction of anisotropic microstructures. To generate anisotropic microstructure samples, the proposed approach necessitates conditional 2D-DGMs that are trained with a dataset of 2D micrographs from three orthogonal planes, thereby enabling conditional sampling from the following estimated distributions:

$$\hat{\mathbf{x}}_{xy} \sim p_\theta(\mathbf{x}_0 | c = c_1) \qquad \text{Eq. 13}$$

$$\hat{\mathbf{x}}_{xz} \sim p_\theta(\mathbf{x}_0 | c = c_2) \qquad \text{Eq. 14}$$

$$\hat{\mathbf{x}}_{yz} \sim p_\theta(\mathbf{x}_0 | c = c_3) \qquad \text{Eq. 15}$$

where $\hat{\mathbf{x}}_{xy}$, $\hat{\mathbf{x}}_{xz}$ and $\hat{\mathbf{x}}_{yz}$ represent the sampled 2D data from the estimated conditional distributions using models $p_\theta(\mathbf{x}_0|c_1)$, $p_\theta(\mathbf{x}_0|c_2)$ and $p_\theta(\mathbf{x}_0|c_3)$, $c$ denotes input condition ($c_1$ or $c_2$ or $c_3$, each corresponding to one of the three orthogonal planes), and $\mathbf{x}_0$ denotes the data combining $\mathbf{x}_{xy}$, $\mathbf{x}_{xz}$ and $\mathbf{x}_{yz}$ (i.e., $\mathbf{x}_{xy} \cup \mathbf{x}_{xz} \cup \mathbf{x}_{yz}$). In particular, $p_\theta(\mathbf{x}_0|c)$ becomes a conditional 2D-DGM for generating 2D samples for a particular plane, which is a model of Markov chain trained with the sequence of $p_\theta(\mathbf{x}_{t-1}|\mathbf{x}_t, c)$ where $t = [T, T-1, \ldots, 2, 1]$. Training this model is straightforward, as it requires only the separate 2D micrographs observed at xy, xz, and yz planes, without concerning for spatial connectivity in 3D space. It should be highlighted that the neural network model utilized for 2D-to-3D reconstruction in this research is $p_\theta(\mathbf{x}_0|c)$. Therefore, both the dataset and the model itself operate within a 2D space.

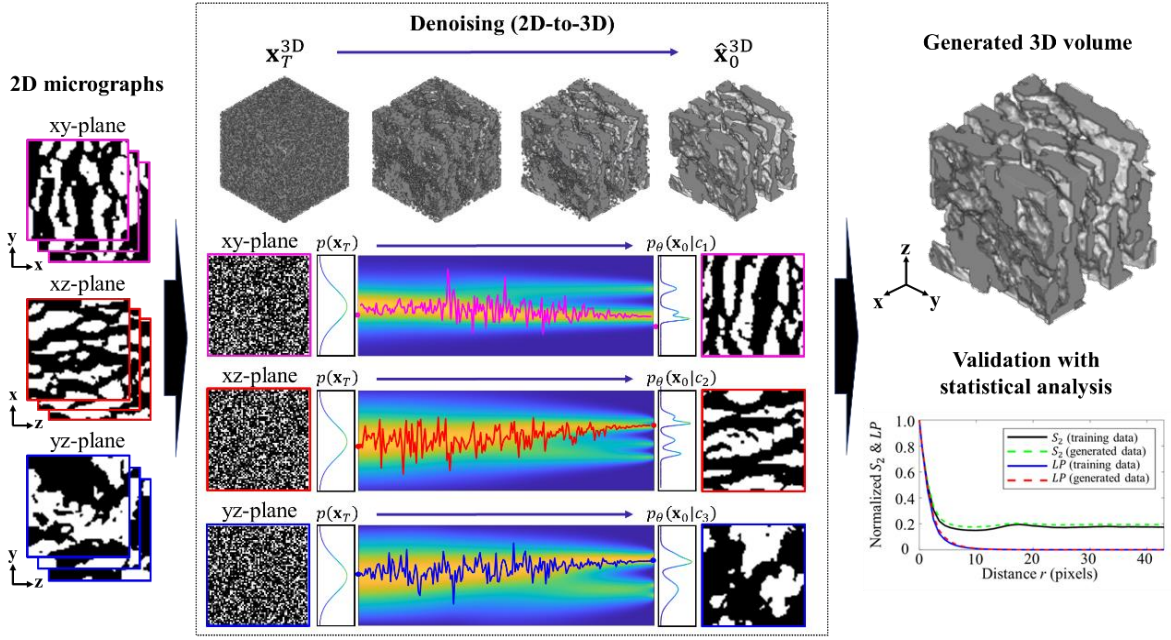

**Figure 3.** Schematic of the proposed framework for conditional denoising diffusion-based generation of 3D anisotropic microstructures (i.e., $\hat{\mathbf{x}}_0^{3D}$) from 2D micrographs.

Utilizing the 2D-DGMs trained solely on the 2D micrographs, this study presents a framework for 2D-to-3D reconstruction of anisotropic microstructures as illustrated in Figure 4. The basic principle of the proposed methodology is to change the target plane (xy or yz or yz), where denoising (i.e., reverse diffusion) with the trained 2D-DGM takes place, during the denoising process for generating a 3D structure. For instance, if we have multiple 2D samples denoised from arbitrary time step $t+1$ and target plane $c$ using the trained 2D-DGM (i.e., $p_\theta(\mathbf{x}_t|\mathbf{x}_{t+1},c)$), the subsequent denoising can take place with $p_\theta(\mathbf{x}_{t-1}|\mathbf{x}_t^*,c_1)$ at xy-plane as shown in Figure 4(a). In addition, $\mathbf{x}_t$ is represented as $\mathbf{x}_t^*$ to note that input is rearranged since the target plane has changed from the previous reverse diffusion step. Next, the target plane changes from xy-plane to xz-plane, so the input condition now should be $c=c_2$, to denoise the 2D samples $\mathbf{x}_{t-1}^*$ with $p_\theta(\mathbf{x}_{t-2}|\mathbf{x}_{t-1}^*,c_2)$ and bring them closer to the actual data $\mathbf{x}_{xz}$. Then, the target plane changes from xz-plane to yz-plane and the sample is denoised again with $p_\theta(\mathbf{x}_{t-3}|\mathbf{x}_{t-2}^*,c_3)$. This process is repeated until the 2D samples are fully denoised to formulate a 3D sample $\hat{\mathbf{x}}_0$. In other words, the proposed framework involves multiple 2D denoising processes occurring simultaneously, conditioned on input condition $c$ while preserving spatial connectivity. The rationale behind the multi-plane denoising process is predicated on the distinct characteristic of DGMs, wherein $\mathbf{x}_t$ (the original input data for the trained 2D-DGM) and $\mathbf{x}_t^*$ (the input data arranged for a different target plane) closely resemble each other due to the thousands of multiple diffusion steps [84]. Therefore, the discrepancy between $\mathbf{x}_t$ and $\mathbf{x}_t^*$ can be assumed to be small enough in a single reverse diffusion step (i.e., $p_\theta(\mathbf{x}_{t-1}|\mathbf{x}_t, c)$) thus, the overlap of multiple 2D denoising processes at different planes can result in the generation of a 3D microstructure (Figure 4).

However, it is worth noting that unlike the original multi-plane denoising diffusion used for 2D-to-3D reconstruction of isotropic microstructures [84], the 2D samples need to be denoised with the models of three different conditional distributions (i.e., $p_\theta(\mathbf{x}_0|c_1)$, $p_\theta(\mathbf{x}_0|c_2)$ and $p_\theta(\mathbf{x}_0|c_3)$) to generate 3D anisotropic microstructures. This could exacerbate the discrepancy between $\mathbf{x}_t$ and $\mathbf{x}_t^*$, potentially leading the model $p_\theta(\mathbf{x}_{t-1}|\mathbf{x}_t^*, c)$ to operate improperly. In other words, since $\mathbf{x}_t^*$ is the data rearranged according to the new target plane, there must be disharmony which could cause the model $p_\theta(\mathbf{x}_{t-1}|\mathbf{x}_t^*, c)$ to deviate significantly from the desired Markov chain model trained with the sequence of $p_\theta(\mathbf{x}_{t-1}|\mathbf{x}_t, c)$ with $t = [T, T-1, \ldots, 2, 1]$. Furthermore, since the distributions of the 2D micrographs on the three orthogonal planes are distinct (i.e., anisotropic), it is more challenging to preserve the spatial connectivity within the anisotropic 3D microstructural geometry.

To address this problem, a modified harmonizing sampling approach (Figure 4(b)) is presented in this study. As the target plane changes with each reverse diffusion step, a harmonizing step can be introduced to reduce disharmony within the sample $\mathbf{x}_t^*$ and to generate a harmonized sample $^h\mathbf{x}_t^*$ that can be written as:

$$p(^h\mathbf{x}_t^*|\mathbf{x}_{t-1}) = N(^h\mathbf{x}_t^*; \sqrt{1-\beta_t}\mathbf{x}_{t-1}, \beta_t\mathbf{I}) \qquad \text{Eq. 16}$$

which is equivalent to the forward process (Eq. 1). This process can be regarded as a process of renoising the sample followed by an additional denoising step (i.e., resampling [84, 87]), which can be repeated for the predefined number of harmonizing steps ($n_h$), to guide the model $p_\theta(\mathbf{x}_{t-1}|\mathbf{x}_t^*, c)$ to operate more closely to $p_\theta(\mathbf{x}_{t-1}|\mathbf{x}_t, c)$ as illustrated in Figure 4(b). In particular, as the denoising diffusion across multiple planes for anisotropic microstructures involves three distinct estimated conditional distributions, maintaining the spatial connectivity that was partially established during the preceding reverse diffusion step is crucial. Thus, the following equation of modified harmonized sampling (i.e., denoising) is utilized for 2D-to-3D reconstruction of anisotropic microstructures:

$$p_\theta(\mathbf{x}_{t-1}|\gamma \cdot {}^h\mathbf{x}_t^* + (1-\gamma) \cdot {}^{h_0}\mathbf{x}_t^*, c) \qquad \text{Eq. 17}$$

where ${}^{h_0}\mathbf{x}_t^*$ is the sample without any harmonizing step applied, and $\gamma$ is a coefficient for blending ${}^h\mathbf{x}_t^*$ and ${}^{h_0}\mathbf{x}_t^*$. In other words, Eq. 17 is for considering the harmonized sample at the current reverse diffusion step and the output from the previous reverse diffusion step (i.e., the output from the harmonized sample at the previous target plane) simultaneously for enhancing spatial connectivity in 3D space. It is again worth noting that Eq. 17 does not represent a new model. The model remains the same as $p_\theta(\mathbf{x}_0|c)$, which is the Markov chain with $p_\theta(\mathbf{x}_{t-1}|\mathbf{x}_t, c)$ for $t = [T, T-1, \ldots, 2, 1]$. The only aspect that has changed for multi-plane denoising diffusion is the input of the model. The generation results and the parametric study regarding the proposed approach are discussed in detail in the subsequent sections (section 3.1).

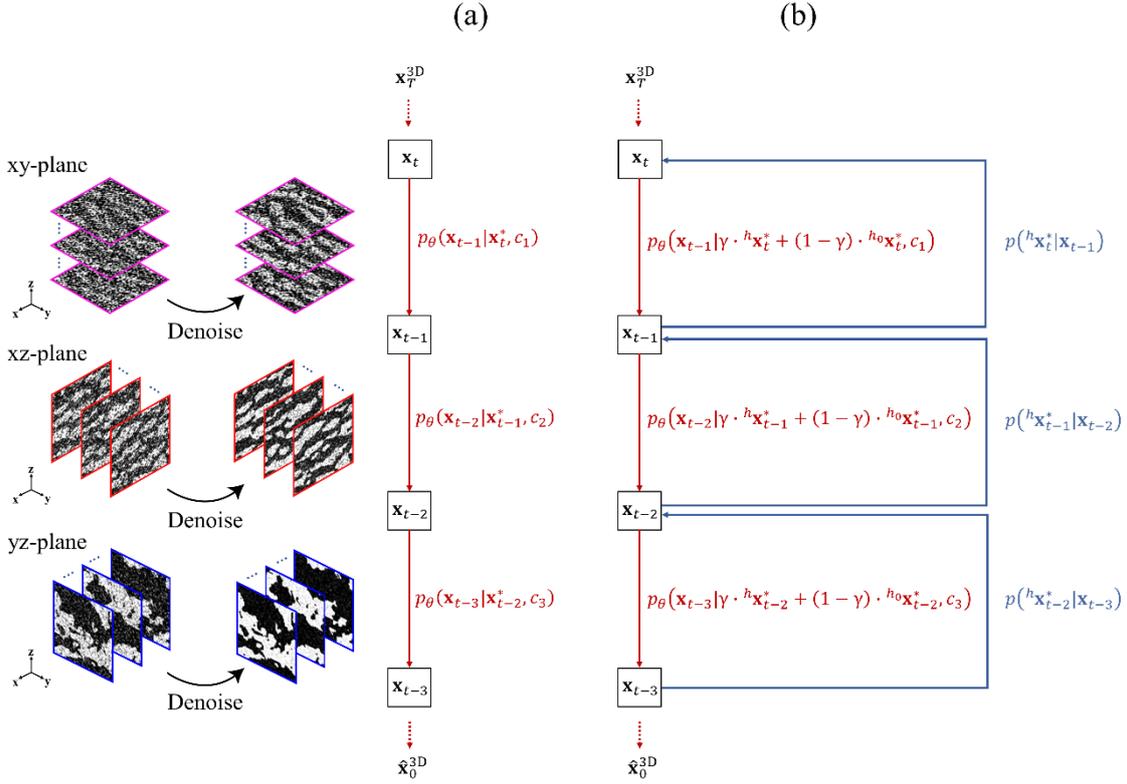

**Figure 4.** Schematic of multi-plane denoising diffusion for generating 3D anisotropic microstructures with conditional 2D-DGMs: (a) vanilla multi-plane denoising diffusion, (b) multi-plane denoising diffusion with modified harmonizing step.

2.4 Implementation details

In order to construct DGMs, the U-Net architecture [88] is adopted for the denoising model $p_\theta(\mathbf{x}_{t-1}|\mathbf{x}_t, c)$. It is worth noting that $p_\theta(\mathbf{x}_{t-1}|\mathbf{x}_t, c)$ can be built with different neural network models. However, U-Net has been used widely for DGMs as it maintains the same dimensions for both the input and the output [68, 71, 72, 80].

In this study, the number of channels for the input of U-Net depends on the number of phases in the material of interest (e.g., 1 for binary microstructure (section 3.1 and 3.2) and 3 for three-phase material (section 3.3)). Once a noised sample $\mathbf{x}_t$ is provided, $\mathbf{x}_t$ is denoised according to the current time step $t$ and condition $c$. In particular, the conditional embedding for $c$, which is used to differentiate the target plane (xy or xz or yz), is pooled and incorporated to the time step embedding, which is formulated using the positional encoding approach [89, 90]. The U-Net models then incorporate conditioning on $c$ through cross attention by concatenating the key-value pairs of self-attention layers and the embeddings [80]. The number of forward/reverse diffusion steps ($T$) is set to be 1000 using the linear noising schedule, with the values of $\beta_t$s evenly spaced between the range of $\beta_1 = 10^{-4}$ and $\beta_T = 10^{-2}$. The loss function (Eq. 6) is optimized utilizing the Adam optimizer [91] with the learning rate of $10^{-4}$. The overall hyperparameters of the proposed DDPMs for 2D-to-3D reconstruction is summarized in Table 1. In addition, the models were trained on Nvidia RTX A6000 graphics processing units (GPU) with a batch size if 32 and 60,000 training steps.

**Table 1.** Hyperparameters used to construct conditional DDPM

| Parameters of conditional DDPM | |
|---|---|
| Number of input channels | 1, 3 |
| Size of input image | 64, 96 |
| Probability of omitting the conditional signal | 0.1 |
| Number of forward/reverse diffusion steps | 1000 |
| Dimension of conditional embedding | 512 |
| Noising schedule | Linear |
| Optimizer | Adam optimizer; learning rate of $10^{-4}$ |
| Batch size | 32 |
| Dimension (U-Net) | 128 |
| Dimension multiples | 1,2,4,8 |
| Layer attentions | False, True, True, True |
| Layer cross-attentions | False, True, True, True |
| Number of Resnet blocks | 3 |
| Number of attention heads | 8 |
| Dimension of attention head | 64 |

2.5 Preparation of datasets

    To validate the proposed methodology for 2D-to-3D reconstruction of microstructures, experimentally obtained 2D micrographs were used for reconstruction of 3D microstructures. The chosen materials for validation include a lithium-ion battery separator [92], carbon fiber

composites [93], and lithium-ion nickel manganese cobalt (NMC) cathode samples [94]. To train conditional DGMs for 2D-to-3D reconstruction, 2D micrographs from three orthogonal planes of each material were randomly sampled from the open-access tomographic data [92-94]. It is worth noting that despite the three-dimensional nature of the dataset, merely a random subset of 2D slices was used to train the 2D-DGMs. For each material category, a subset of 300 images was sampled from each orthogonal plane. The dataset was then augmented by incorporating either vertical or horizontal flips to the original images, without rotation, to preserve the anisotropy of the microstructural geometry.

2.6 Quantitative evaluation metrics

To validate the performance of the 2D-to-3D microstructure reconstruction methods, the generated samples must be evaluated using appropriate quantitative metrics. Since the two-point correlation function [46, 47, 57, 59, 60, 95] is the most widely used metric for evaluating the reconstruction results, it was used as one of the evaluation metrics in this study. The equation for obtaining the two-point correlation function $f_c^{S_2}$ can be expressed as:

$$f_c^{S_2}(\mathbf{p}_1, \mathbf{p}_2) = S(\mathbf{p}_1)S(\mathbf{p}_2) \qquad \text{Eq. 18}$$

where $S$ is a function that turns to 1 if a material phase of interest is present at given locations $\mathbf{p}_1$ and $\mathbf{p}_2$, and 0 otherwise. However, according to Torquato [96], relying solely on the two-point correlation function might be inadequate for assessing the statistical equivalence of microstructure samples. Thus, the lineal-path correlation function $f_c^{L_p}$ [60, 97] was also employed to evaluate the statistical equivalence of microstructure samples which can be written as:

$$f_c^{L_p}(\mathbf{p}_1, \mathbf{p}_2) =$$
$$= \begin{cases} 1 & \text{if a straight line from } \mathbf{p}_1 \text{ to } \mathbf{p}_2 \text{ lies within material phase of interest} \\ 0 & \text{otherwise} \end{cases}$$

Eq. 19

In addition, for a more precise evaluation of the reconstructed microstructures, it is essential to analyze the material behavior in relation to the geometry of the microstructures. To further validate the proposed 2D-to-3D microstructure reconstruction method, the material properties of reconstructed 3D microstructure samples, including those of fiber-reinforced composites and battery electrode materials, were analyzed through computational simulations [2, 98]. For the case of fiber-reinforced composites, the macroscopic anisotropic mechanical properties (e.g., elastic moduli, Poisson ratios and shear moduli) of the generated 3D microstructure samples were analyzed using the computational homogenization method [2, 99]. The detailed procedure for the computational homogenization is explained in Appendix A. Furthermore, the relative diffusivity values of the generated 3D microstructures of NMC cathodes were also computed in 3D space using Taufactor [98], an open-source software developed for analyzing the microstructures of materials such as batteries and fuel cell materials. Using the evaluation metrics introduced, the statistical equivalence of the 2D-to-3D reconstructed microstructure samples was assessed in terms of both spatial distributions and physical material properties. This assessment serves the quantitative validation of the proposed framework, which will be detailed in the following sections.

## 3. Results

3.1 Generated 3D microstructure samples of battery separator

To demonstrate the capabilities of the proposed 2D-to-3D microstructure reconstruction framework, the anisotropic 3D microstructure samples ($64 \times 64 \times 64$) of lithium-ion battery separator were generated as shown in Figure 5. As shown in the figure, realistic 3D microstructure samples (see Figure 5(a)) were generated with 2D-DGMs while preserving spatial connectivity. Furthermore, when compared to the original 2D slices at three different orthogonal planes (as seen in Figure 5(b)), the 2D slices of the generated anisotropic 3D microstructure samples (Figure 5(c)) exhibit similar microstructural morphologies. It is worth noting that no additional optimization was applied to ensure morphological similarity, which highlights the capability of the DGM-based 2D-to-3D reconstruction of morphologically similar anisotropic structures. Furthermore, the two-point correlation functions and the lineal path functions for the 2D slices of the generated 3D samples from three orthogonal planes were computed and compared with those of the original 2D micrographs, as shown Figure 5(d). While $f_C^{S_2}$ and $f_C^{L_p}$ of the 2D slices in the generated 3D samples at xy and xz-planes closely match the original 2D micrographs, there are relatively larger discrepancies at the yz-plane, probably due to the complex geometry with fewer patterns in the 2D micrographs at the yz-plane. Overall, the results indicate that the generated samples well align with the original 2D micrographs in terms of spatial and morphological characteristics.

Furthermore, multiple generation processes were executed with varying $\gamma$ and $n_h$ to analyze their effects on the quality of the generated samples as shown in Figure 6 and Figure 7. Figure 6 displays the relative errors between the correlation functions (i.e., $f_c^{S_2}$ and $f_c^{L_p}$) of 2D slices in the generated 3D samples and the training images. As can be seen in the figure, the generated samples rapidly become closer to the training images in terms of $f_C^{S_2}$, as $\gamma$ increases from 0.15 to 0.5. This suggests that the use of harmonizing steps is effective in the 2D-to-3D reconstruction of anisotropic microstructures with 2D-DGMs. Then, the error increases as $\gamma$ increase to 1. This emphasizes the importance of using $\gamma$ to blend samples with and without harmonizing steps to reconstruct anisotropic microstructures. Meanwhile, the impact of varying $\gamma$ diminishes as the number of harmonizing steps increases. This might be explained by the reduced effect of blending the samples as the harmonizing step is repeatedly applied, which is equivalent to repeated sampling with the model $p_\theta(\mathbf{x}_{t-1}|\mathbf{x}_t^*, c)$. Figure 6 also indicates that the discrepancy decreases as $n_h$ increases. However, it is important to note that the increase of $n_h$ leads to longer computational times, as it requires conducting more forward (Eq. 1) and reverse (Eq. 5) passes. Thus, the authors suggest that finding an appropriate value for $n_h$ is necessary to ensure that the sampling process computationally affordable. On the other hand, the relative error of $f_C^{L_p}$ does not exhibit a clear trend compared to $f_C^{S_2}$. In addition, Li et al. [100] showed that the two-point correlation function is more effective in capturing the morphological details of microstructure compared to the lineal path function. Applications of MCR with two-point correlation functions are also known to perform better compared to those utilizing the lineal path function in previous studies [46, 49, 100]. Therefore, putting more emphasis on $f_C^{S_2}$ and taking computation efficiency into account, the 2D-to-3D reconstruction results for the remainder of this paper were conducted with $n_h = 4$ and $\gamma = 0.5$. Results of an ablation study of the model used for 2D-to-3D reconstruction are also provided in Appendix B.

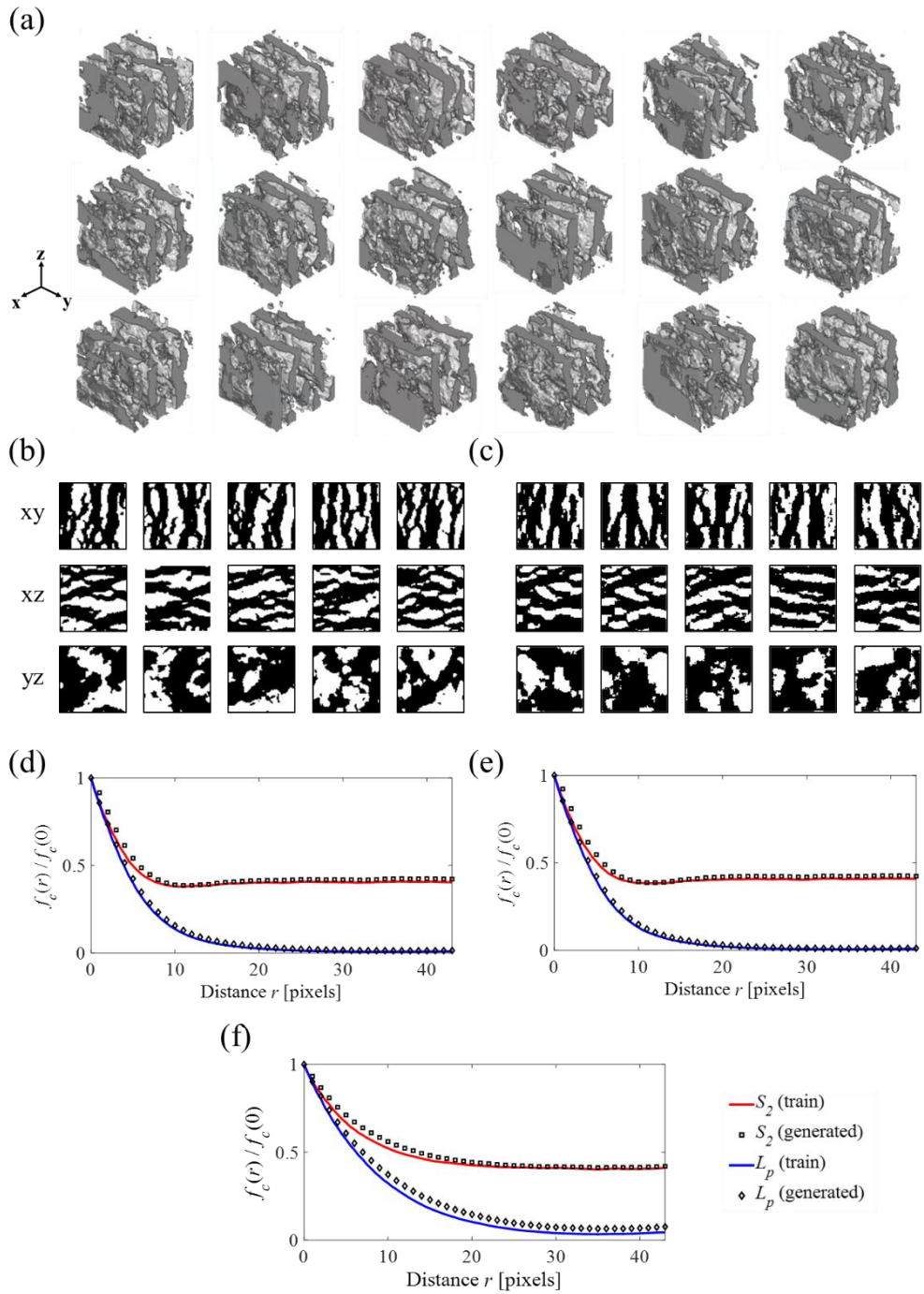

**Figure 5.** Generated anisotropic microstructure (battery separator) samples with the proposed 2D-to-3D DGM: (a) 3D volumes of the generated samples, (b) 2D samples for training the conditional DGM, (c) 2D slice views of the generated 3D samples, (d) spatial correlation functions in the xy-plane, (e) spatial correlation functions in the xz-plane, and (f) spatial correlation functions in the yz-plane.

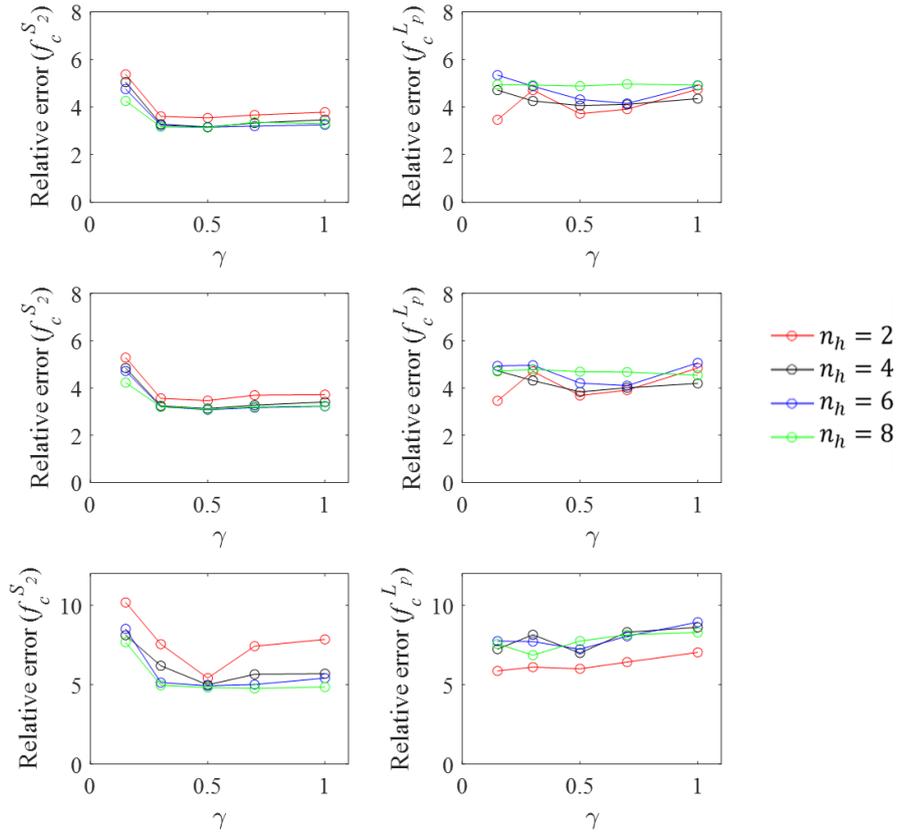

**Figure 6.** Variation of reconstruction performance with different values of $\gamma$ and $n_h$: relative errors of $f_c^{S_2}$ and $f_c^{L_p}$ at three orthogonal planes

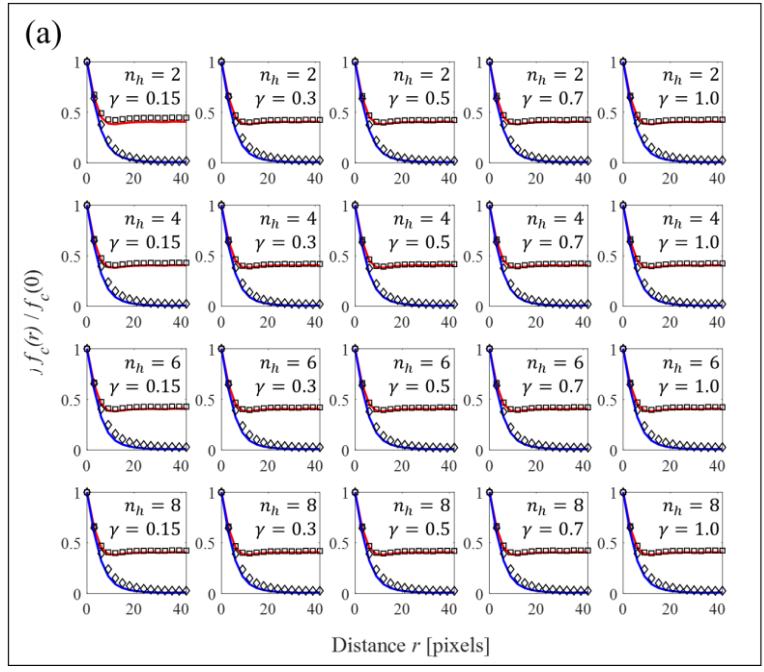

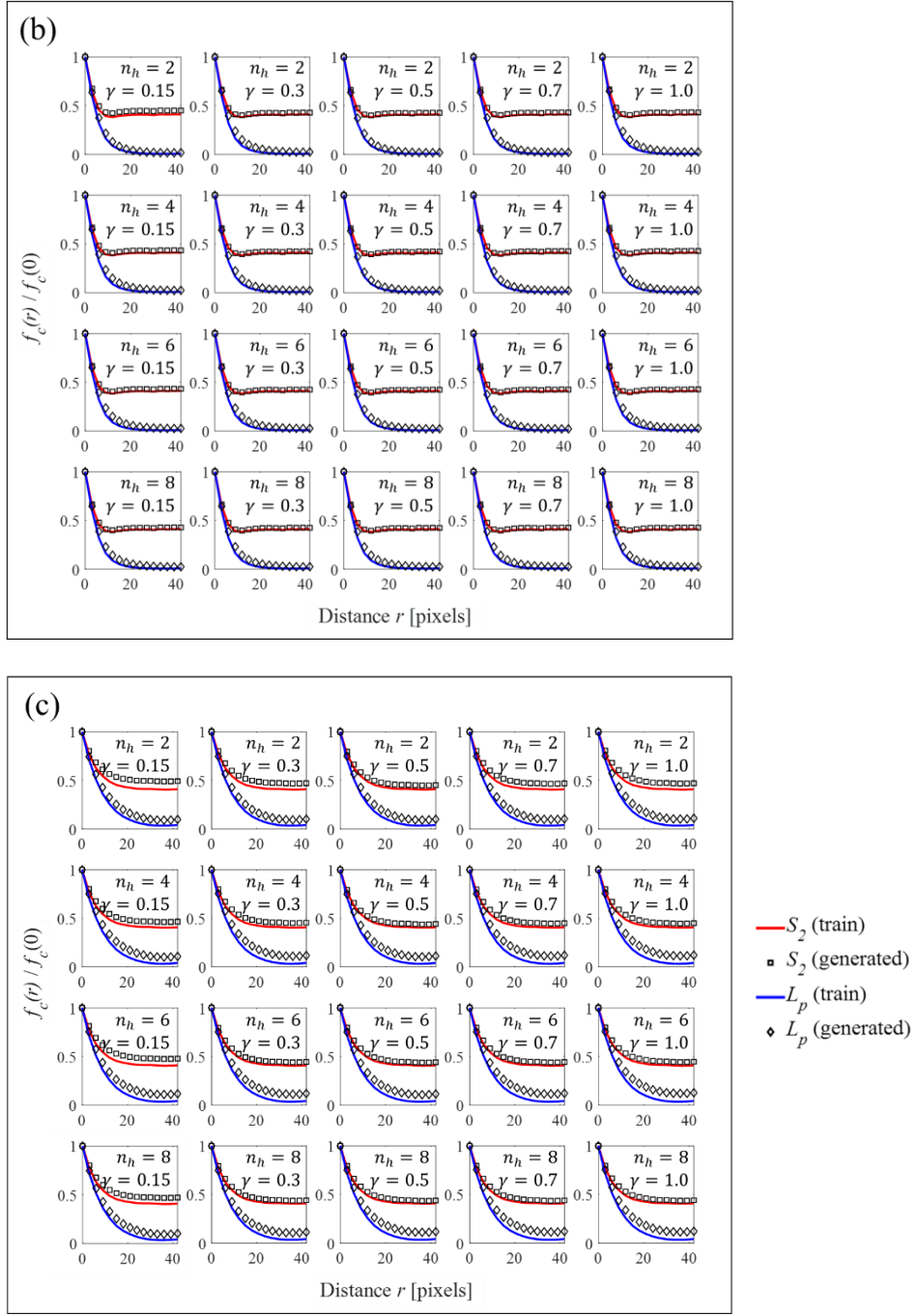

**Figure 7.** Variation of reconstruction performance with different values of $\gamma$ and $n_h$: (a) $f_c^{S_2}$ and $f_c^{L_p}$ at xy-plane, (b) $f_c^{S_2}$ and $f_c^{L_p}$ at xz-plane and (c) $f_c^{S_2}$ and $f_c^{L_p}$ at yz-plane.

3.2 Generated 3D microstructure samples of fiber-reinforced composites

Figure 8 illustrates the generated 3D samples of carbon fiber composites ($64 \times 64 \times 64$) from 2D micrographs, along with the comparison of the spatial correlation functions from the generated and training samples. As shown in Figure 8(a) and Figure 8(c), the spherical morphology of the fiber was well reproduced at the xy-plane. The spatial correlation functions of the 2D slices in the generated 3D samples also show good agreement with the training micrographs Figure 8(d)-(f). The 2D slices at the xz and yz planes also suggest that the fiber alignment is well reproduced, at least upon visual examination. In addition, since the fiber orientation distribution significantly affects the resultant material behavior, [39, 99, 101], it is important to reproduce equivalent orientation distribution in the case of fiber-reinforced composites. Moreover, the characteristics of fibers (e.g., orientation distribution and volume fraction) significantly influence the elastic behavior of composites [99, 101]. In other words, the material properties of generated samples may differ from those of the training samples, even if the morphology of fibers is well reproduced.

In this regard, the anisotropic mechanical properties, such as elastic moduli, Poisson ratios, shear moduli and coefficient of thermal expansion (CTE) of the samples, were computed based on the computational homogenization (refer to Appendix A) to validate whether the generated samples have material properties equivalent to the original samples. From the original tomographic data [93], 50 different volumetric subsamples that have same dimensions ($64 \times 64 \times 64$) with the 2D-to-3D reconstructed samples were randomly sampled. The randomly sampled microstructures were then treated as representative volume elements (RVEs) to analyze the macroscopic material properties using FEA-based computational homogenization. Figure 9 and Figure 10 illustrate the statistical analysis results for the anisotropic material properties of both the generated and original samples. The results demonstrate that the material properties were reasonably well reproduced, with the medians aligning closely with those of the training dataset. This result is impressive, considering that no additional constraints or conditions were imposed on the model (i.e., 2D-DGM) to reproduce the material properties in 3D space. Although the deviations are not identical, the results suggest that the generated samples could potentially be considered equivalent to the ground truth (i.e., original tomographic data) in terms of physical material behavior.

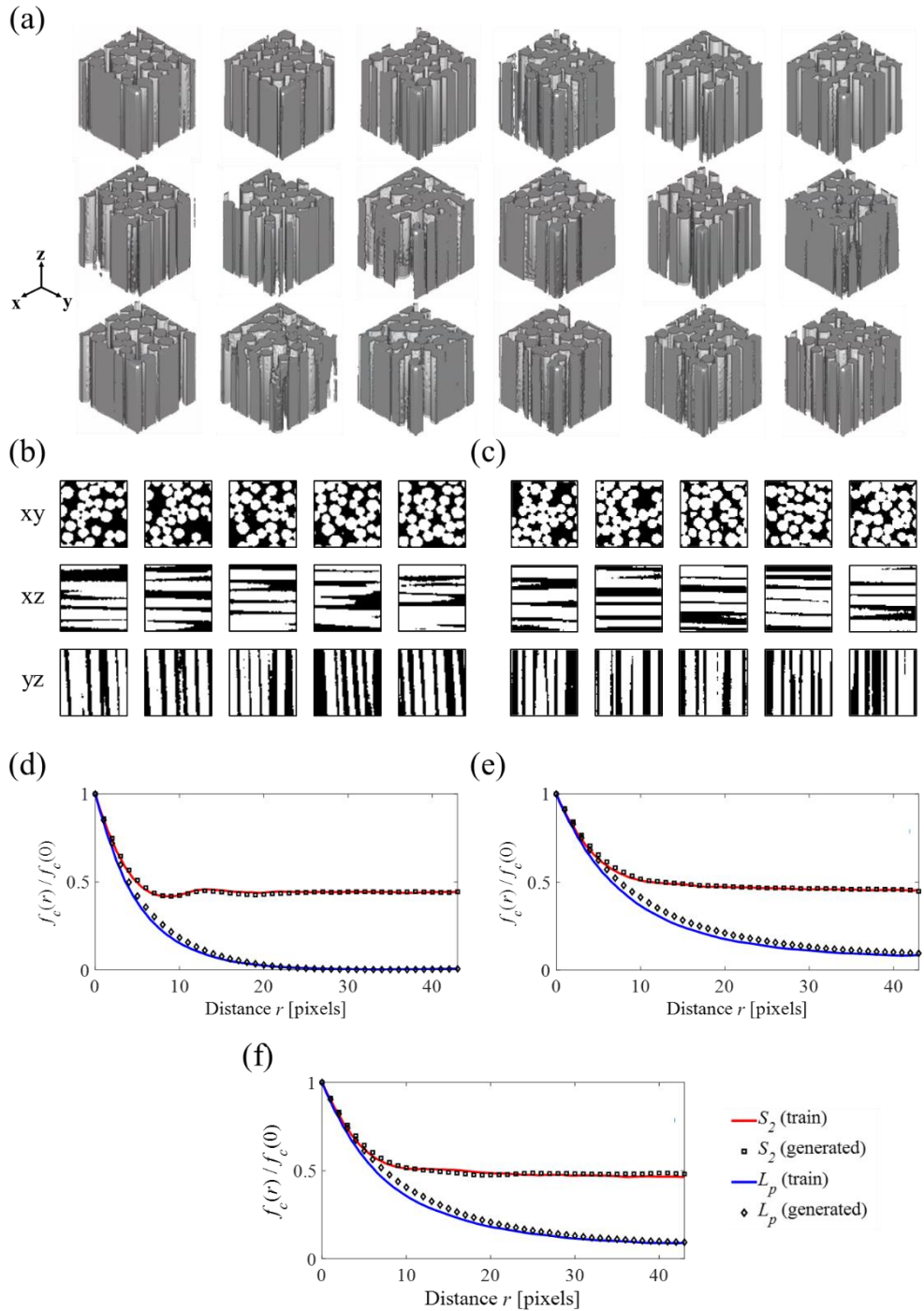

**Figure 8.** Generated anisotropic microstructure (fiber-reinforced composites) samples with the proposed 2D-to-3D DGM: (a) 3D volumes of the generated samples, (b) 2D samples for training the conditional DGM, (c) 2D slice views of the generated 3D samples, (d) spatial correlation functions in the xy-plane, (e) spatial correlation functions in the xz-plane, (f) spatial correlation functions in the yz-plane.

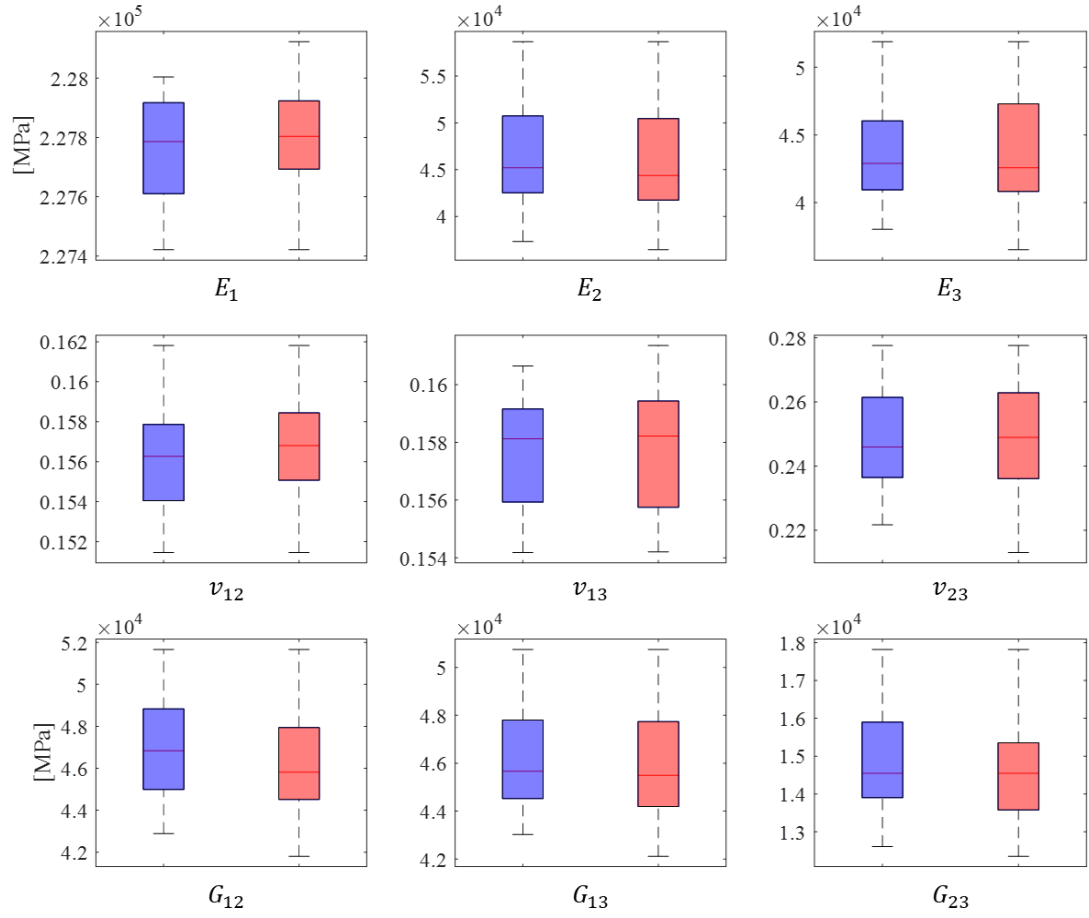

**Figure 9.** Statistical analysis results for anisotropic mechanical properties (elastic moduli ($E_1$, $E_2$ and $E_3$), Poisson ratios ($v_{12}$, $v_{13}$ and $v_{23}$) and shear moduli ($G_{12}$, $G_{13}$ and $G_{23}$)) of ground truth data (blue) and generated(red) microstructure samples of fiber-reinforced composite.

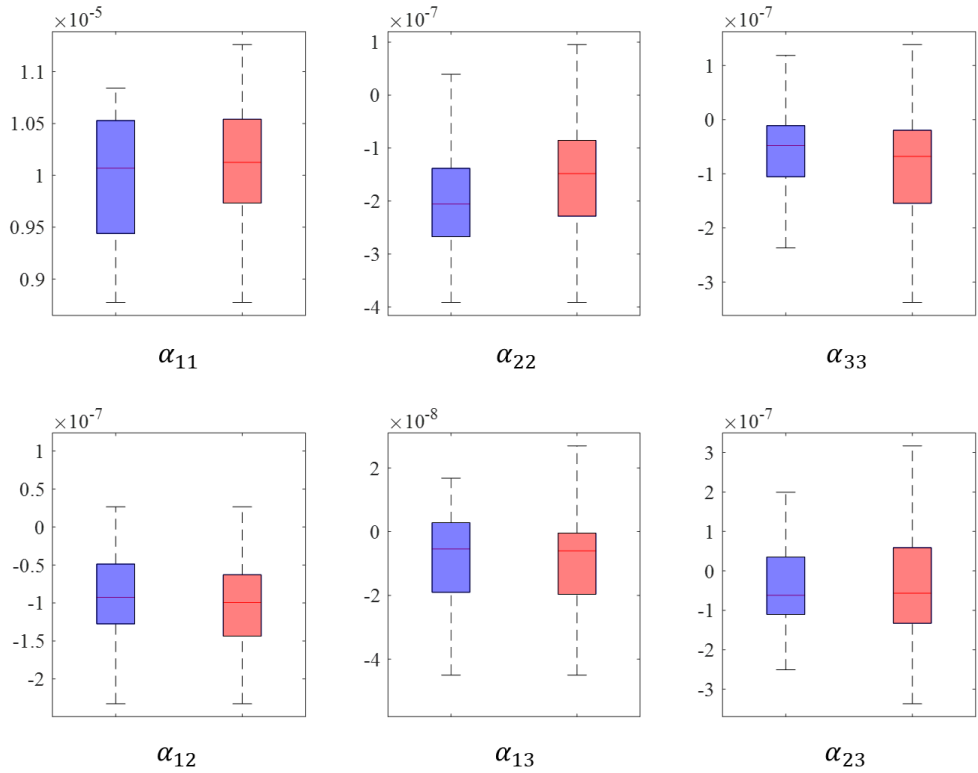

**Figure 10.** Statistical analysis results for CTE properties ($\alpha_{11}$, $\alpha_{22}$, $\alpha_{33}$, $\alpha_{12}$, $\alpha_{13}$ and $\alpha_{23}$) of ground truth data (blue) and generated (red) microstructure samples of fiber-reinforced composite.

3.3 Generated 3D microstructure samples of battery electrode material

Since there are many materials consist of several phases (i.e., multi-phase) more than two, reconstruction of microstructures within these materials is an important topic in the field of materials engineering. In particular, reconstructing microstructures of three-phase battery electrode materials is gaining significant attention, as their complex microstructures are closely related to mechanical and electrochemical performance as well as mass transport [94, 102]. In this regard, utilizing randomly sampled 2D three-phase NMC cathode microstructure samples, which consist of three phases (pore, active material, and binder), 3D volumes were generated with the proposed 2D-to-3D reconstruction framework, as shown in Figure 11(a). Compared to the original 2D slices (Figure 11(b)), the slices of the generated 3D samples exhibit similar morphologies (e.g., the binder materials attached to the active materials). The spatial correlation functions of the generated samples for each material phase (as shown in Figure 11(d-f)) also closely match those of the training dataset. This implies that the spatial distributions of the three phases were well reproduced in 3D space, even though only 2D micrographs were used to train the 2D-DGM. Figure 12(a) also shows that the volume fraction of each phase was well reproduced, with closely matching medians and deviations.

To validate the results in terms of material properties, the effective relative diffusivity [98] of each material phase is computed using randomly subsampled 3D volumes from the original tomographic data [94], following a similar approach to that in section 3.2. As can be seen in Figure 12(b), the relative diffusivities of 2D-to-3D reconstructed samples demonstrate a high degree of concordance with the ground truth. Given that the effective relative diffusivity was evaluated in 3D space, it is noteworthy that the proposed DGM-based 2D-to-3D reconstruction framework can reproduce this property solely based on the 2D micrographs. This highlights the capability of the framework in capturing not only the morphological and spatial correlations, but also the properties which could contribute to the design of battery electrode materials for future energy storage technologies.

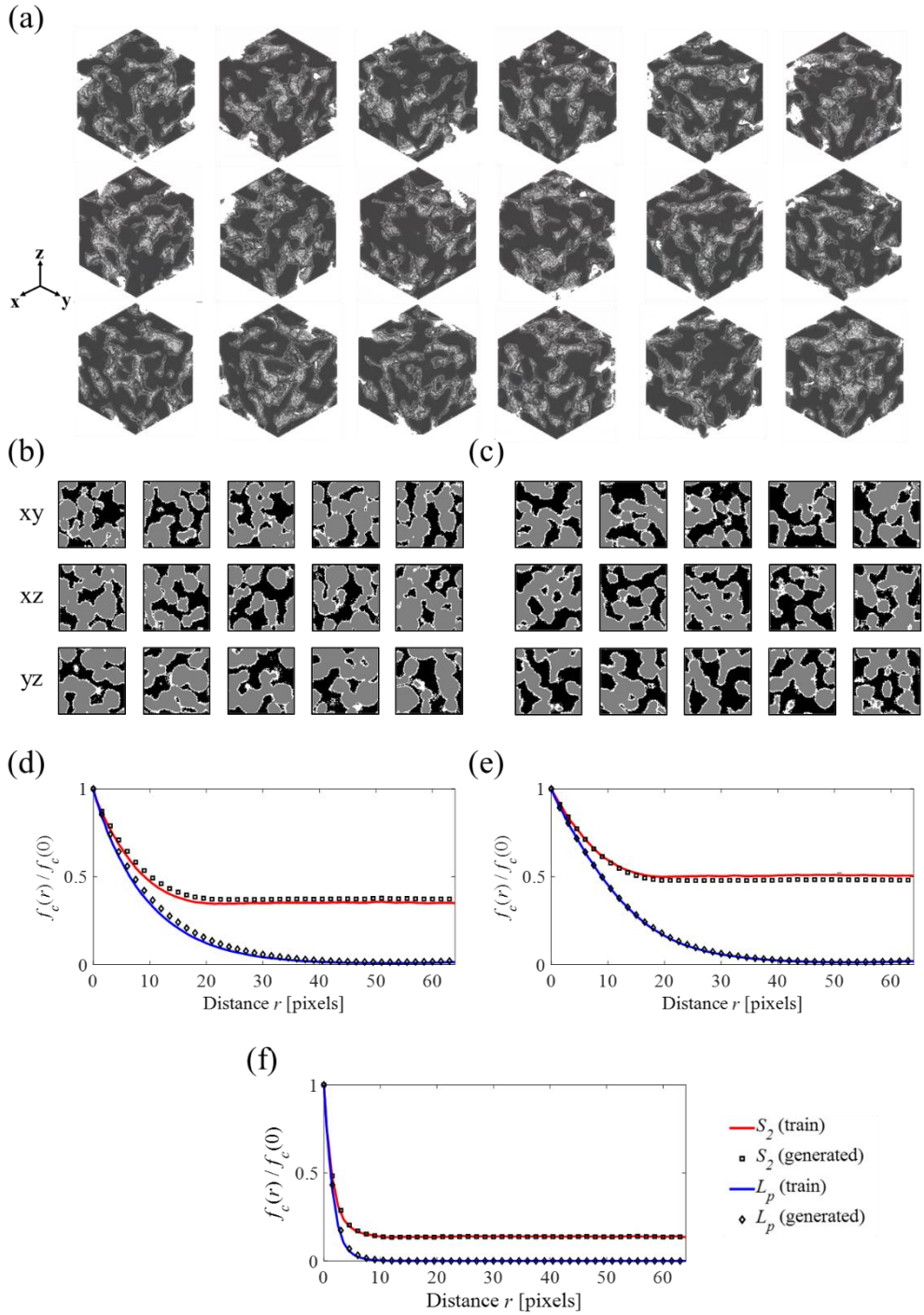

**Figure 11.** Generated multiphase microstructure (battery electrode material) samples with the proposed 2D-to-3D DGM: (a) 3D volumes of the generated samples, (b) 2D samples for training the conditional DGM, (c) 2D slice views of the generated 3D samples, (d) spatial correlation functions of the first phase (pore), (e) spatial correlation functions of the second phase (active material) and (f) spatial correlation functions of the third phase (binder).

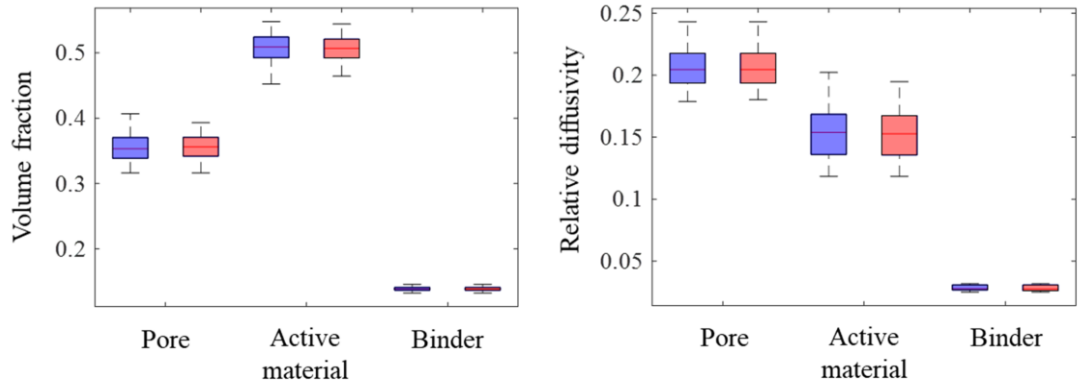

**Figure 12.** Statistical analysis results for properties (volume fraction and relative diffusivity) of ground truth data (blue) and generated (red) microstructure samples of three-phase battery electrode material.

## 4. Discussion

4.1 Model performance

In this study, a DGM-based framework is presented for reconstructing anisotropic microstructures based on conditional DGMs with a guidance signal and multi-plane denoising diffusion using the modified harmonized sampling. By utilizing 2D micrographs sampled from experimental tomographic data, the microstructures of different types of materials, including fiber-reinforced composites, battery separators, and the three-phase microstructures of battery electrode materials, were reconstructed in 3D space. The generated samples exhibit spatial correlation functions that closely match those of the training dataset, which indicates that the spatial distributions of material phases and morphological characteristics were well reproduced. Furthermore, it is noteworthy that the material properties of the generated microstructures within 3D space (e.g., anisotropic mechanical properties and diffusivities), are statistically similar to those of the ground truth 3D tomographic data. Given that the models receive no additional conditions or information about the material behavior beforehand, this performance is impressive which suggests the potential use of the proposed framework for analyzing material properties. This could aid in establishing a process-microstructure-property linkage for specific material systems, as well as in developing the ICME framework. Furthermore, it is worth noting again that since DGMs have outperformed previous generative models (e.g., GANs), without suffering from unstable training and mode collapse issues as reported in recent studies [70-72], the development of DGM-based applications is encouraged in the field of data-driven MCR. In this context, a more sophisticated study aimed at modifying and scaling the proposed DGM-based framework to achieve better sample quality will be undertaken in the future.

4.2 Computational efficiency

Since DGMs involve multiple diffusion steps for forward and reverse processes, the forward pass with the 2D-DGMs (i.e., $p_\theta(\mathbf{x}_{t-1} \mid \mathbf{x}_t)$) should be executed $T$ times. Therefore, it takes several minutes to reconstruct 3D samples with the trained 2D-DGMs – for instance, approximately 7 minutes for generating a $64 \times 64 \times 64$ sample and 16 minutes for a $96 \times 96 \times 96$ sample in this study. However, considering that some of the latest MCR methods even with GPU-aided iterative optimization require several minutes to an hour of computation time to generate a 3D sample comparable to the original micrograph [46, 49], the computation cost of the proposed method can be considered affordable. In addition, several studies have been conducted to reduce the sampling time of DGMs, such as denoising diffusion implicit models (DDIMs) [103] which employ a deterministic version of the sampling equation to shorten the sampling trajectory. These methods could potentially be employed to accelerate the 2D-to-3D reconstruction process. However, since the principle of multi-plane denoising diffusion with the modified harmonized sampling (as discussed in section 2.3) is based on multiple diffusion steps, any study concerning the modification of the sampling trajectory may require careful consideration.

4.3  Rationale for utilizing diffusion-based generative models

As previously mentioned, the proposed 2D-to-3D reconstruction method (section 2.3) utilizes a unique characteristic of DGMs, which is the gradual sampling process for transforming noise into the data structure. This enables the multi-plane denoising diffusion, enabling execution of spatially connected multiple 2D reverse diffusion processes to formulate a 3D sample. Thus, it would be difficult to apply the same approach with other generative models, such as GANs and VAEs, which generate a sample in a single forward pass. Furthermore, it is noteworthy that the proposed methodology allows for anisotropic 2D-to-3D reconstruction using 2D-DGMs trained with 2D images, achieved by merely manipulating the latent variables and without the need for additional models or renderers. Another important contribution of this study is the extension of DGM's capability for dimensionality expansion, applicable not only to isotropic structures but also to anisotropic structures. Future advancements in 2D-DGMs for enhanced 2D-to-3D reconstruction are also anticipated, which may involve incorporating sophisticated model architectures such as the use of transformer encoders to assist in conditioning the models.

4.4  Universality and transferability

The validation results (section [3](#)) demonstrate that 2D-to-3D reconstruction of different types of material systems is available using the proposed DGM-based framework, without any descriptors for characterizing the target microstructures. This implies that 2D-to-3D reconstruction is feasible using an available 2D microstructure dataset, without the need for specialized prior knowledge of microstructural morphology and spatial distributions. However, if a more complex microstructure is considered and the number of material phases increases, a more extensive experimental dataset will be required. In this regard, research on the transferability of DGMs for 2D-to-3D reconstruction should be conducted to address the potential issue of data scarcity. For instance, a 2D-DGM trained for one specific material system could facilitate the 2D-to-3D microstructure reconstruction of other materials.

4.5 Future work

In summary, the scope of this study is focused on proposing and validating a concept for generating anisotropic 3D microstructures from 2D micrographs using conditional 2D-DGMs. For future research directions, scaling and enhancing model architectures to improve the quality of 2D-to-3D reconstructed samples is encouraged. One promising approach involves the incorporation of a pre-trained transformer encoder [90, 104], a powerful tool known for its effectiveness in capturing complex patterns and dependencies in data. Utilizing such an encoder for conditioning the models could substantially refine the accuracy and detail of the generated microstructures. Furthermore, accelerating the sampling process in the proposed DGM-based 2D-to-3D reconstruction to reduce computational costs is also suggested for future research. However, careful consideration is necessary when manipulating the sampling process, as improper adjustments could cause the models to deviate significantly from the desired sampling trajectory. Additionally, exploring the transferability of 2D-DGMs for the reconstruction of microstructures could be a promising direction for future studies. This could help address potential issues of data scarcity and enhance the performance of the models.

Since establishing a well-organized dataset of 3D microstructure samples is an important step in analyzing the physical behavior of particular materials, the proposed framework can aid in constructing a process-microstructure-property linkage within the context ICME. Although it may present challenges, the inverse design of microstructures to achieve desired material properties, based on the microstructure-property linkage supported by the proposed framework, could facilitate high-throughput material development in the future.

## 5. Appendix

5.1 A. Computational homogenization for evaluation of reconstructed microstructure samples

In order to obtain the anisotropic material properties of the 2D-to-3D reconstructed microstructures of fiber-reinforced composites, the FEA-based computational homogenization method, which is based on the asymptotic expansion theory [8, 18], was utilized. Considering the two different domains, which are the macroscale spatial domain $\mathbf{\Omega}$ and the microscale RVE domain $\mathbf{\Theta}$, the gradient of material response can be written as follows:

$$f_{,x_i}^{\varepsilon} = \frac{\partial f}{\partial x_i} + \frac{1}{\varepsilon}\frac{\partial f}{\partial y_i} \qquad \text{Eq(A). 1}$$

$$f_{,y_i}^{\varepsilon} = \frac{\partial f}{\partial y_i} \qquad \text{Eq(A). 2}$$

where $\varepsilon^{-1}$ is the magnification factor with the coordinates $\mathbf{y} = \{y_1, y_2, y_3\}$ (in $\mathbf{\Omega}$) and $\mathbf{z} = \{z_i, z_j, z_k\}$ (in $\mathbf{\Theta}$) have the following relationship:

$$\mathbf{y} = \frac{\mathbf{z}}{\varepsilon} \qquad \text{Eq(A). 3}$$

Based on the above definitions, the governing equations for mechanical problems can be written as follows:

$$\sigma_{ij,y_j}^{\varepsilon} + b_i = 0 \quad \text{in } \mathbf{\Omega}^{\varepsilon} \qquad \text{Eq(A). 4}$$

$$u_i^\varepsilon = \bar{u}_i \quad \text{on} \quad \partial\Omega^\varepsilon \qquad \text{Eq(A). 5}$$

$$\sigma_{ij}^\varepsilon n_j = \bar{t}_i \quad \text{on} \quad \partial\Omega^\varepsilon \qquad \text{Eq(A). 6}$$

where $\sigma_{ij}$ represents the stress tensor, $b_i$ is the body force, $u_i$ is the displacement, $n_j$ is the normal vector, and $\bar{t}_i$ represents the traction. In addition, the superscript $\varepsilon$ is for indicating that the quantity is within the asymptotically expanded domain $\Omega^\varepsilon$. The displacement can be asymptotically expanded as follows:

$$u_i^\varepsilon(\mathbf{y}, \mathbf{z}) = u_i^{(0)}(\mathbf{y}, \mathbf{z}) + \varepsilon u_i^{(1)}(\mathbf{y}, \mathbf{z}) + \varepsilon^2 u_i^{(2)}(\mathbf{y}, \mathbf{z}) + O(\varepsilon^3) \qquad \text{Eq(A). 7}$$

The governing equation Eq(A). 4 can be rewritten incorporating the constitutive law and the thermal expansion as follows:

$$\int_\Omega C_{ijkl} \left[\frac{\partial u_k^\varepsilon}{\partial y_l} - \alpha_{kl}\Delta T\right] \frac{\partial v_i}{\partial y_j} d\Omega = \int_\Omega b_i v_i d\Omega + \int_\Gamma t_i v_i d\Gamma \qquad \text{Eq(A). 8}$$

where $C_{ijkl}$ denotes the elastic stiffness tensor, $v_i$ is the virtual displacement, and $\Delta T$ is the temperature difference. Based on the asymptotically expanded displacement Eq(A). 7, the homogenized elastic stiffness tensor $C_{ijkl}^H$ can be obtained the following equation:

$$C_{ijkl}^H(\mathbf{y}) = \frac{1}{|\Theta|} \int_\Theta C_{ijkl}(\mathbf{y}, \mathbf{z}) \left(\frac{\partial \chi_p^{kl}}{\partial z_q} + I_{pqkl}\right) d\Theta \qquad \text{Eq(A). 9}$$

$$\frac{\partial}{\partial z_j}\left(C_{ijkl}\left(I_{klmn}+\frac{\partial \chi_k^{mn}}{\partial z_l}\right)\right)=0 \qquad \text{Eq(A). 10}$$

$$I_{klmn}=0.5(\delta_{mk}\delta_{nl}+\delta_{nk}\delta_{ml}) \qquad \text{Eq(A). 11}$$

where $\chi_k^{mn}$ is the displacement influence function [8, 21, 99]. To obtain homogenized CTE, the volumetric-averaged strain with unit thermal loading (i.e., $\Delta T=1$) was obtained using the following equation:

$$\alpha_{kl}^H(\mathbf{y})=\frac{1}{|\Theta|}\int_\Theta \varepsilon_{kl}(\mathbf{y},\mathbf{z})d\Theta \qquad \text{Eq(A). 12}$$

In addition, the periodic boundary condition (PBC) was applied with the constraint equations separately applied for the nodes on the faces (F), edges (E) and vertices (V) of the RVEs (Figure A 1) to avoid over-constraint. The equations for PBC are as

$$\begin{aligned}u_i^{F2}-u_i^{F1}&=L_x\varepsilon_{i1};\\ u_i^{F4}-u_i^{F3}&=L_y\varepsilon_{i2};\\ u_i^{F6}-u_i^{F5}&=L_z\varepsilon_{i3}\end{aligned} \qquad \text{Eq(A). 13}$$

$$\begin{aligned}u_i^{E2}-u_i^{E4}&=L_x\varepsilon_{i1}+L_y\varepsilon_{i2};\\ u_i^{E1}-u_i^{E3}&=L_x\varepsilon_{i1}-L_y\varepsilon_{i2};\\ u_i^{E6}-u_i^{E8}&=L_x\varepsilon_{i1}+L_z\varepsilon_{i3};\\ u_i^{E5}-u_i^{E7}&=L_x\varepsilon_{i1}-L_z\varepsilon_{i3};\\ u_i^{E11}-u_i^{E9}&=L_y\varepsilon_{i2}+L_z\varepsilon_{i3};\end{aligned} \qquad \text{Eq(A). 14}$$

$$u_i^{E10} - u_i^{E12} = L_y \varepsilon_{i2} - L_z \varepsilon_{i3}$$

$$u_i^{V3} - u_i^{V5} = L_x \varepsilon_{i1} + L_y \varepsilon_{i2} + L_z \varepsilon_{i3};$$

$$u_i^{V2} - u_i^{V8} = L_x \varepsilon_{i1} + L_y \varepsilon_{i2} - L_z \varepsilon_{i3};$$

$$u_i^{V7} - u_i^{V1} = -L_x \varepsilon_{i1} + L_y \varepsilon_{i2} + L_z \varepsilon_{i3};$$

$$u_i^{V4} - u_i^{V6} = L_x \varepsilon_{i1} - L_y \varepsilon_{i2} + L_z \varepsilon_{i3}$$

Eq(A). 15

with the correlation lengths $L_x$, $L_y$ and $L_z$. The exemplary simulation results (i.e., equivalent stress fields) with applied PBC for obtaining anisotropic mechanical properties with an RVE are shown in Figure A 2. The mechanical properties of the fiber and matrix used for the validation are as shown in Table A 1.

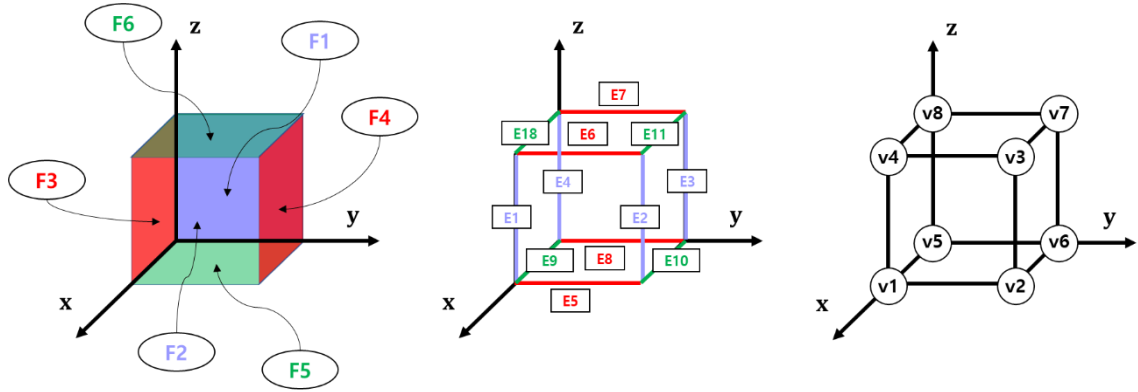

**Figure A 1.** Set of nodes on the faces, edges, and vertices for applying PBC without over-constraint.

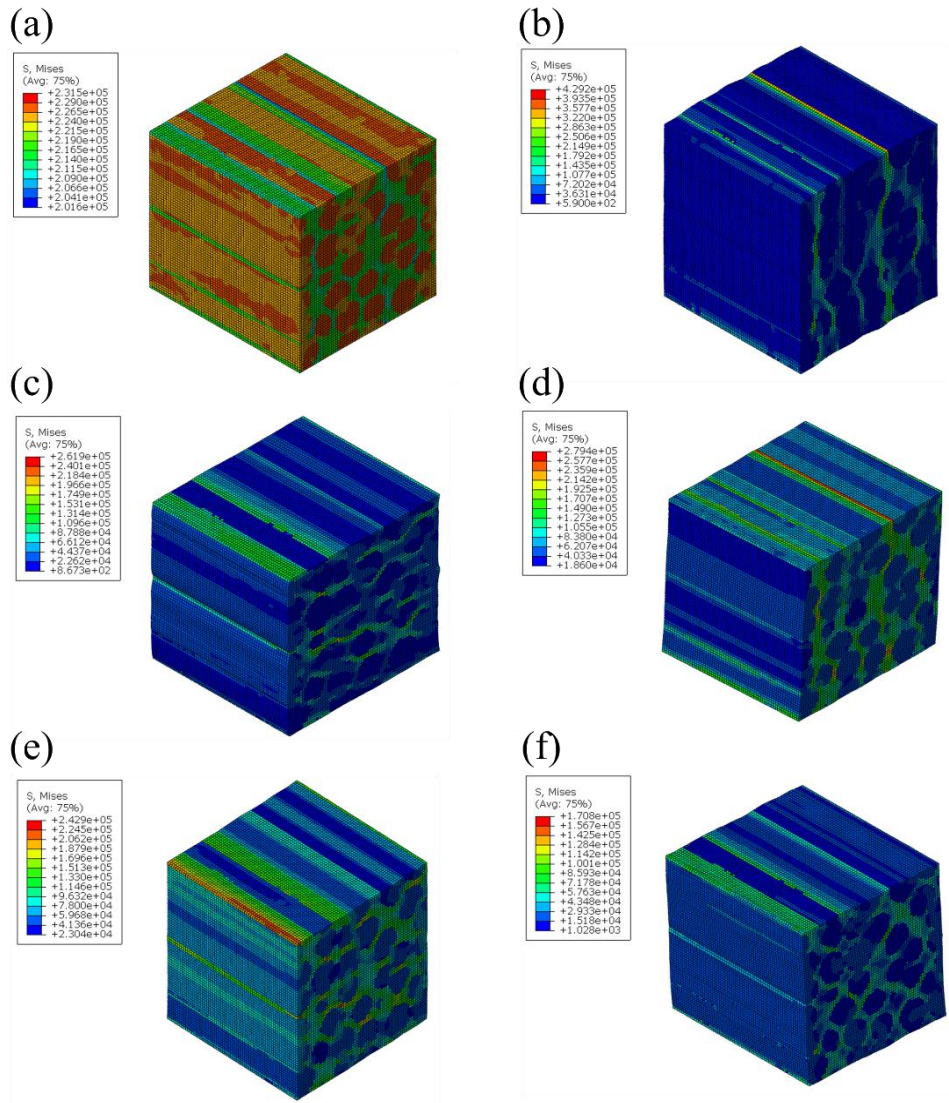

**Figure A 2.** Exemplary simulation results for obtaining anisotropic mechanical properties of fiber-reinforced composites with PBF and directional loading: (a) 11-direction, (b) 22-direction, (c) 33-direction, (d) 12-direction, (e) 13-direction and (f) 23-direction

**Table A 1.** Mechanical properties of fiber and matrix used for computational homogenization [99, 105].

|  | Fiber | Matrix |
|---|---|---|
| $E_1$[GPa] | 230 | 225 |
| $E_2, E_3$[GPa] | 15 |  |
| $G_{12}$[GPa] | 24 |  |
| $G_{23}$[GPa] | 5.03 |  |
| $v_{12}$ | 0.2 | 0.14 |
| $v_{23}$ | 0.25 |  |
| $\alpha_1 [10^{-6}/K]$ [ | 18 | 0.164 |
| $\alpha_2, \alpha_3 [10^{-6}/K]$ | 1.2 | 0.164 |

5.2  B. Ablation study of DGMs for generating microstructure samples

Figure B 1 shows the results of an ablation study on the conditional 2D-DGM used for the 2D-to-3D reconstruction of microstructures in lithium-ion battery separators. For each type of model architecture, the relative errors of the spatial correlation functions are computed and averaged after 10 different random seed initializations. As can be seen in the figure, the relative error decreases as the model dimensions (i.e., number of base channels in U-Net) increase and with the increased number of ResNet blocks and attention heads. There are relatively larger errors for the case of the yz-plane, likely due to the more complex geometry with fewer patterns in the 2D micrographs compared to those at the xy and xz-planes (section 3.1).

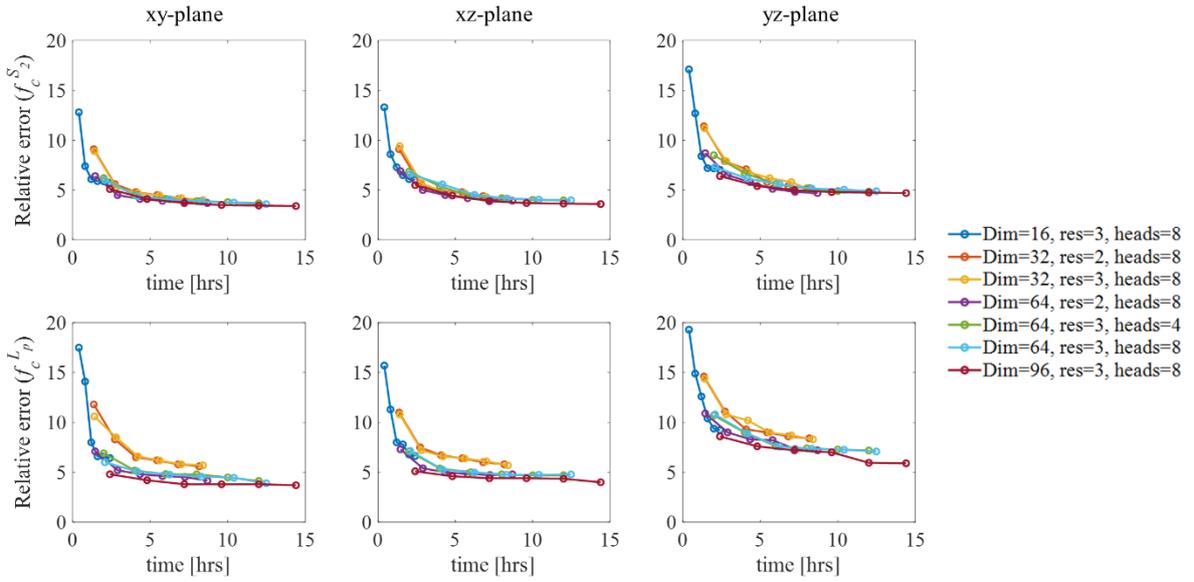

**Figure B 1.** Ablation study for model (i.e., 2D-DGM) architecture with the evaluation metrics ($f_c^{S_2}$ and $f_c^{L_p}$)